\documentclass[journal=jacsat,manuscript=article]{achemso}

\usepackage[T1]{fontenc} 

\author{Kazutomo Kawaguchi}
\affiliation
{Institute of Science and Engineering, Kanazawa University, Kanazawa, Ishikawa 920-1192, Japan}
\email{kkawa@wriron1.s.kanazawa-u.ac.jp}
\author{Hidemi Nagao}
\affiliation
{Institute of Science and Engineering, Kanazawa University, Kanazawa, Ishikawa 920-1192, Japan}
\author{Hideo Shindou}
\affiliation
{Department of Lipid Life Science, National Center for Global Health and Medicine, Shinjuku-ku, Tokyo 162-8655, Japan}
\alsoaffiliation
{Department of Medical Lipid Science, Graduate School of Medicine, University of Tokyo, Bunkyo-ku, Tokyo 113-0033, Japan}
\author{Hiroshi Noguchi}
\affiliation
{Institute for Solid State Physics, University of Tokyo, Kashiwa, Chiba 277-8581, Japan}

\title[An \textsf{achemso} demo]
  {Conformations of three types of ultra-long-chain fatty acids in multi-component lipid bilayers}

\abbreviations{IR,NMR,UV}

\begin{document}


\begin{abstract}
Ultra-long-chain fatty acids (ULCFAs) are biosynthesized in certain types of tissue,
but their biological roles remain unknown.
Here, we report how the conformation of ULCFAs depends on the length and unsaturated-bond ratio
of the ultra-long chains and the composition of the host bilayer membrane using  molecular dynamics simulations.
The ultra-long chain of ULCFAs flips between the two leaflets and fluctuates among three conformations:
elongated, L-shaped, and turned.
Further, we found that the saturated ultra-long chain
exhibited an elongated conformation more frequently
than the unsaturated chain.
In addition,
the truncation of the ultra-long chain at C26 had little effect on the remaining ULCFAs.
ULCFAs respond to lipid-density differences in the two leaflets,
and the ratio of the elongated and turned conformations changed to reduce this difference.
However,
in cholesterol-containing membranes,
ULCFAs behave as with no density difference  after the flip--flop of cholesterol removes the difference. 
\end{abstract}

\section{Introduction}
Phospholipids have a polar head group and two hydrocarbon tails (fatty acids)
 and are the most abundant lipids in the membranes of living cells. 
Phospholipids more than 1000 molecular species are biosynthesized
 by a combination of a head group and two hydrocarbon tails.
Phospholipids have a range of
 structural and functional roles in cells\cite{meer08, shin09, anto15}
 and are
 biosynthesized through two pathways:
 the Kennedy pathway (de novo pathway) \cite{kennedy56} and the Lands' cycle (remodeling pathway) \cite{lands58}.
In the Kennedy pathway, glycerol-3-phosphate (G3P) acyltransferases (GPATs) use G3P and acyl-CoA
 as substrates to produce lysophosphatidic acid (LPA).
LPA is converted to phosphatidic acid (PA) through LPAAT reaction,
 which is also a lysophospholipid acyltransferase (LPLAT) reaction.
PA is then further converted to other classes of phospholipids.
Phospholipids with all types of polar head groups are newly biosynthesized during the Kennedy pathway.
Next, in the Lands' cycle, fatty acids of phospholipids are changed by the concerted actions of phospholipase (PLA)1/2s and LPLATs.
Thus, the LPLAT reaction contributes to generating membrane phospholipid diversity
 through the Kennedy pathway and the Lands' cycle \cite{val22}.

The ratio of unsaturated bond in hydrocarbon tails varies widely in biomembranes.
With increase in the ratio of unsaturated bonds in lipid membranes,
 the fluid--gel transition temperature decreases,
 therefore, the unsaturated bond ratio is important for controlling the fluidity of membranes. \cite{koyn98}
In cells, membranes are laterally and vertically heterogeneous.
Laterally, a microdomain called lipid raft is formed as a platform of protein receptors and their effectors
and consists of saturated phospholipids, cholesterol, and others. \cite{simo00,ling10}
The two leaflets of bilayer membranes have different lipid compositions. \cite{dale03}
For example, the outer leaflet of the red blood cell membrane mainly consists of phosphatidylcholine (PC) and sphingomyelin (Sph),
 whereas the inner leaflet mainly consists of phosphatidylethanolamine (PE) and phosphatidylserine (PS). \cite{verkleij73}

The phospholipid composition of membranes affects their biological functions.
Recently, several LPLAT-knock-out (KO) mice have been reported to show changes in functions.
Further, a deficiency of LPLAT3 (AGPAT3), which produces docosahexaenoic acid (DHA) containing phospholipids \cite{koeb10}
 showed visual dysfunction \cite{shin17a} and male infertility \cite{iizuka17}.
Arachidonic acid containing PC is produced by LPLAT12 (LPCAT3) \cite{hishi08}.
LPLAT12-KO mice showed drastic reductions in membrane arachidonate levels,
 which was neonatally lethal due ton extensive triacylglycerol (TG) accumulation and dysfunction in enterocytes \cite{hashi15}.
Phospholipids also work as mediators.
Platelet-activating factor (PAF) is a potent phospholipid mediator produced by LPLAT9 (LPCAT2) \cite{shin07}, and
LPLAT9-KO mice with low levels of PAF showed a relief from neuropathic pain \cite{shin17b}.

Each phospholipid tail typically consists of 14-22 carbon atoms.
Fatty acids composed of around C22 are called very long-chain fatty acids (VLCFAs).
However, longer chains composed of 32-36 carbon atoms with six double bonds
 at the sn-1 position of phosphatidylcholine (PC) have been discovered \cite{sang05, mcma09, bara13}
 and are called ultra-long-chain fatty acids (ULCFAs).
In contrast,
 intermediate lengths (C26--C28) of fatty acids and saturated ultra-long chains have not been found in living cells.
The physicochemical properties of these ULCFAs with respect to their structures
 (length, large number of unsaturated bonds, and sn-1 position),
 as well as
the biological roles of ULCFA containing phospholipids, are 
 not fully understood. 
For example,
 the elongase of very long chain fatty acid-4 (ELOVL4) is one of biosynthetic enzymes of ULCFA-CoA
 from DHA-CoA and its KO mice showed slightly reduced visual function \cite{hark12}. 
ULCFAs have also been reported to be lipid mediator precursors \cite{jun17}. 
However,
an LPLAT enzyme producing ULCFAs containing phospholipids has also not been identified.

All-atom molecular dynamics (MD) simulations have been widely applied
 to study lipid membranes \cite{muel06, vent06, nogu09, vena15, marr19}
 and can reproduce membrane properties well.
All-atom MD simulations for VLCFAs, in which the tail length is maximally 24,
 have also been performed \cite{ramo16, gupt16, rog16, mann17, wang18}
 and have shown that the long acyl chain is interdigitated into the opposite leaflet.
In our previous study \cite{kk20}, one ULCFA molecule
 (dotriacontahexaenoic acid (C32:6) containing phosphatidylcholine (dTSPC, C32:6-C18:0)) 
 embedded in a single-component phospholipid membrane has been simulated using all-atom MD simulations.
We have clarified that the long tail of the ULCFA flips between the two leaflets
 and that the ULCFA can respond to the density difference between the two leaflets and respond to the density difference change.
However, the dependence of these conformational changes on the ULCFA structures has not yet been investigated.

In this study, we examined the conformation of three types of ULCFAs in a lipid bilayer
 using all-atom MD simulations.
 The ULCFAs under study were dTSPC (C32:6-C18:0),
 hexacosatetraenoic acid (C26:4) containing phosphatidylcholine (HSPC, C26:4-C18:0), and
 lacceroic acid (C32:0) containing phosphatidylcholine (LSPC, C32:0-C18:0).
The chain length and the number of unsaturated bonds varied.
The membranes containing these ULCFAs were analyzed using membrane phospholipids containing
 distearoyl PC (DSPC, C18:0-C18:0) and stearoyl-DHA PC (SDPC, C18:0-C22:6).
An asymmetric membrane, in which one leaflet consisted of a DPSC and the other consisted of SDPC,
was examined. The ratio of lipids in the two leaflets varied.
Moreover, we describe the effects of the flip--flop of cholesterol (CHOL),
 which compensates for the difference in lipid density between the upper and lower leaflets.

\section{Methods}

\subsection{ULCFAs}
The phospholipids used in this study are shown in \ref{fig1}.
A single ULCFA molecule (dTSPC (C32:6-C18:0), HSPC (C26:4-C18:0), or LSPC (C32:0-C18:0)) was inserted into the lipid bilayer.
dTSPC has an ultra-long chain of 32 carbons with six double bonds at the sn-1 position, as shown in \ref{fig1}(c).
HSPC has a hexacosatetraenoic acid chain with four double bonds at the sn-1 position (\ref{fig1}(d)).
LSPC has a lacceroic acid chain with no double bonds at the sn-1 position, as shown in \ref{fig1}(e).
All of the ULCDAs have a stearoyl chain at the sn-2 position.
HSPC is constructed by truncating the long hydrocarbon chain of dTSPC at C26,
and LSPC is by the saturation of the long chain of dTSPC.

\subsection{Asymmetric Bilayer}
\begin{figure}
  \includegraphics[width=140mm]{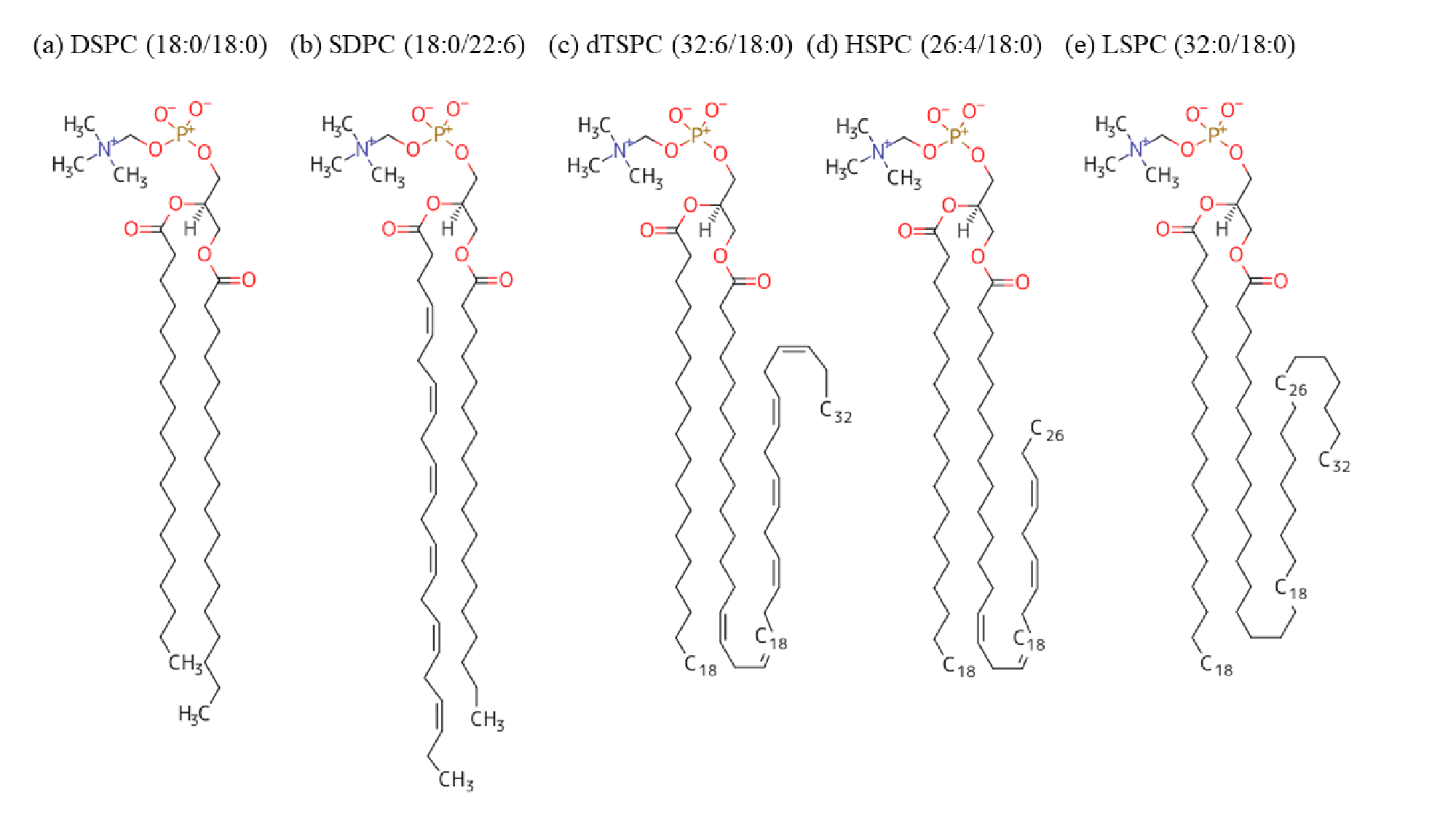}
  \caption{Molecular structures of (a) DSPC, (b) SDPC, (c) dTSPC, (d) HSPC, and (e) LSPC.}
  \label{fig1}
\end{figure}

In this study, we considered a DSPC/SDPC mixture asymmetric bilayer as a host lipid bilayer.
DSPC (C18:0-C18:0) contains two saturated stearoyl chains with no double bonds, as shown in \ref{fig1}(a).
SDPC (C18:0-C22:6) contains a stearoyl chain at the sn-1 position
and a docosahexaenoyl chain with six double bonds at the sn-2 position, as shown in \ref{fig1}(b).
We prepared six types of host lipid bilayers
with different numbers of lipids, as shown in \ref{t1}.
A ULCFA molecule was inserted into the upper leaflet.

\begin{table}
  \caption{Host lipid bilayers for three types of ULCFAs used in the present MD simulations.}
  \label{t1}
  \begin{tabular}{lll}
    \hline
    Membrane  & Upper leaflet & Lower leaflet \\
    \hline
    DS89-SD100 & 89 DSPCs & 100 SDPCs \\
    DS99-SD90  & 99 DSPCs &  90 SDPCs \\
    DS99-SD98  & 99 DSPCs &  98 SDPCs \\
    SD89-DS100 & 89 SDPCs & 100 DSPCs  \\
    SD99-DS90  & 99 SDPCs &  90 DSPCs  \\
    SD99-DS98  & 99 SDPCs &  98 DSPCs  \\
    \hline
  \end{tabular}
\end{table}

\subsection{Asymmetric Bilayer with Cholesterol Molecules}
To investigate the effects of the flip--flop of cholesterol (CHOL) on the conformation of the ULCFAs,
 we added cholesterol to both the upper and lower leaflets which consist of DSPCs or SDPCs.
The numbers of molecules in each membrane bilayer are listed in \ref{t2}.

\begin{table}
  \caption{Asymmetric bilayers containing cholesterol molecules used for MD simulations.}
  \label{t2}
  \begin{tabular}{lll}
    \hline
    Membrane  & Upper leaflet & Lower leaflet \\
    \hline
    DSPC/dTSPC/CHOL20 & 20 CHOLs, 89 DSPCs, 1 dTSPC & 20 CHOLs, 100 DSPCs \\
    DSPC/dTSPC/CHOL40 & 40 CHOLs, 89 DSPCs, 1 dTSPC & 40 CHOLs, 100 DSPCs \\
    SDPC/dTSPC/CHOL20 & 20 CHOLs, 89 SDPCs, 1 dTSPC & 20 CHOLs, 100 SDPCs  \\
    SDPC/dTSPC/CHOL40 & 40 CHOLs, 89 SDPCs, 1 dTSPC & 40 CHOLs, 100 SDPCs  \\
    \hline
  \end{tabular}
\end{table}

\subsection{Molecular Dynamics Simulations}
The CHARMM 36 force field \cite{charmm36} and TIP3P water model \cite{tip3p} were used 
 to model lipid and water molecules, respectively.
The system temperature and pressure were controlled at 343 K using the Nos{\'e}--Hoover thermostat \cite{nh}
 and 0.101 MPa using a semi-isotropic Parrinello-Rahman barostat \cite{pn}, respectively.
The membranes have zero surface tension.
Calculation of the van der Waals interaction was truncated with a cutoff radius of 1-1.2 nm using a switching scheme.
The electrostatic interactions were calculated using the particle mesh Ewald method \cite{pme}.
The time step for the numerical calculation of the equations of motion was 2.0 fs.
The membranes were equilibrated for 200\,ns, followed by 800\,ns production runs
(1\,$\mu$s and $3$\,$\mu$s for DSPC/CHOL and SDPC/CHOL, respectively).
All MD simulations were performed using GROMACS version 2020.4 \cite{gromacs}.
Initial configurations were generated using Membrane Builder on the CHARMM-GUI website \cite{jo2008charmm, lee2015charmm}.
Visual Molecular Dynamics (VMD) software \cite{vmd} was used for visualization of molecular images.
The center of the lipid bilayer along the bilayer normal direction was used as the origin of the vertical ($z$) coordinates.

To calculate the lipid area in the tensionless membranes,
 we performed  MD simulations for a pure symmetric membrane consisting of 100 DSPCs or 100 SDPCs per leaflet.
From these MD simulations, we obtained that the area per lipid of DSPC and SDPC are 0.63 and 0.74 nm$^2$, respectively.
Thus, SDPCs have a slightly larger area than DSPCs because of their unsaturated bonds.

\section{Results and Discussion}

\subsection{Comparison of Three Types of ULCFAs: Conformation of the sn-1 chain of ULCFAs}
\begin{figure}
  \includegraphics[width=150mm]{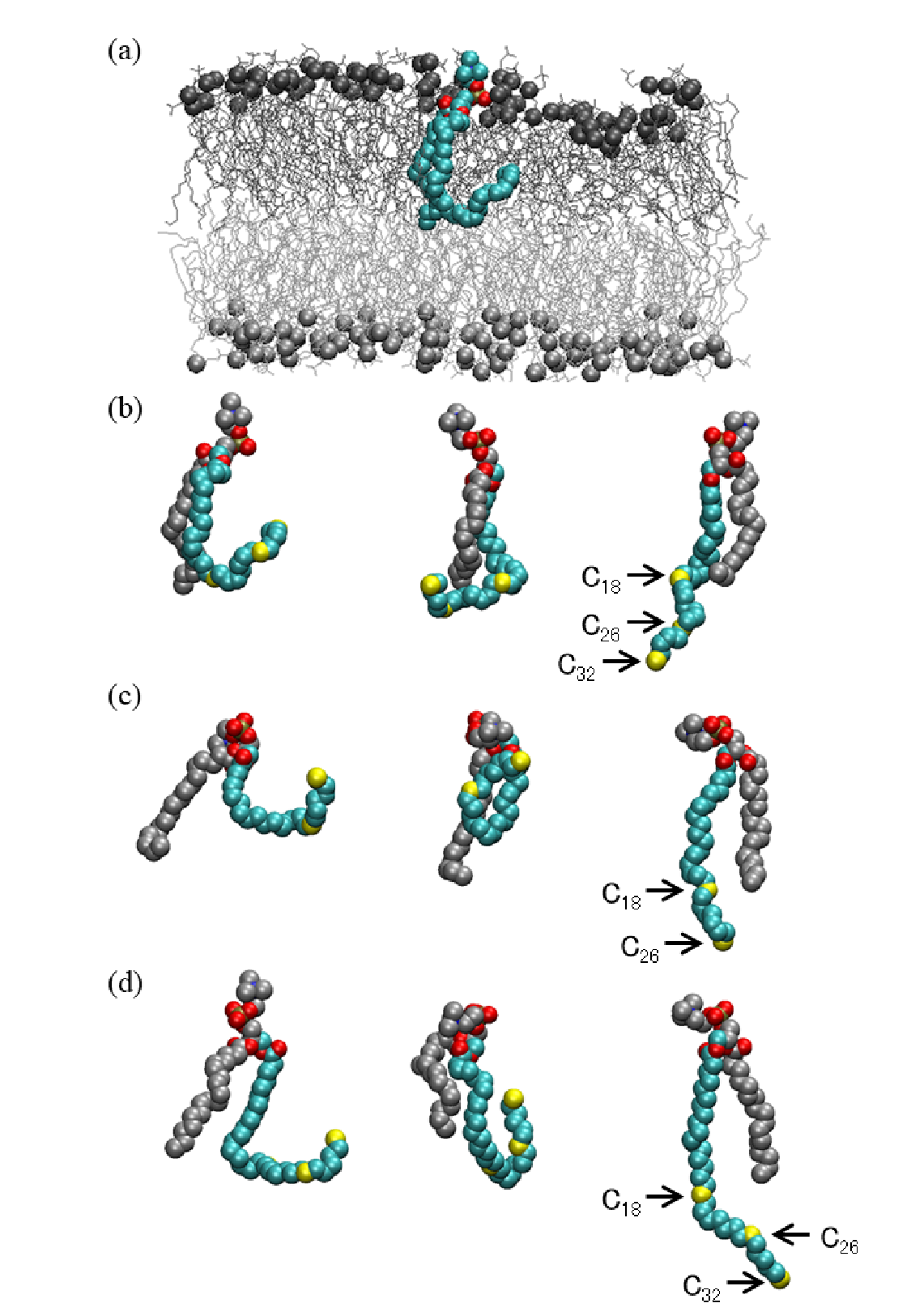}
  \caption{Snapshots of ULCFA. (a) dTSPC in the DS89-SD100 membrane.
           dTSPC is represented by colored spheres by atom type (C, cyan; N, blue; O, red; P, gold).
           DSPC and SDPC molecules are shown in dark and light gray, respectively.
           Dark and light gray spheres represent the phosphate atoms of DSPC and SDPC, respectively.
           (b)--(d) Examples of the conformations of (b) dTSPC, (c) HSPC, and (d) LSPC.
           Carbon atoms in the sn-1 chain are shown in cyan and other carbon atoms are shown in gray.
           Hydrogen atoms are not shown for clarity.
           }
  \label{fig2}
\end{figure}

First, we considered the three types of ULCFAs in asymmetric membranes in the absence of cholesterol (\ref{t1}).
The conformations of ULCFAs are shown in \ref{fig2} and Movie S1.
Large conformational changes were observed in the sn-1 chain of ULCFAs, as reported in our previous study~\cite{kk20}.
The sn-1 chain of the ULCFAs fluctuates between the upper and lower leaflets
 and exhibits three types of conformation:  L-like, turned, and elongated,
 as shown in the left, middle, and right panels, respectively, of \ref{fig2}(b)--(d).
In \ref{fig2}(a), an L-shaped conformation of dTSPC is shown with the other lipids.
Here, the C$_{3}$-C$_{18}$-C$_{32}$ angle is 52$^{\circ}$, and the C$_{32}$ atom is located in the upper leaflet (the $z$-coordinate of C$_{32}$ is $0.77$\,nm).
In the elongated conformation, the sn-1 chain was deep in the lower (opposite) leaflet.
All three types of ULCFAs exhibited these three conformations under all simulated conditions:
 however, their ratios were dependent on differences in the types and densities of the lipid molecules
 in the upper and lower leaflets.

\begin{table}
\caption{Transit times (ns) of C$_{26}$ and C$_{32}$ atoms through the lipid bilayer.
           Standard deviations are described in parentheses.}
\label{t3}
\begin{tabular}{llrrrr}
\hline
ULCFA &lipid       & C$_{26}$ &       & C$_{32}$ &       \\
\hline
dTSPC & ds89-sd100 &  6.0 & (3.2) &  8.9 & (7.7) \\
      & ds99-sd90  &  7.7 & (5.7) &  7.4 & (5.6) \\
      & ds99-sd98  &  6.1 & (3.4) & 14.9 & (9.1) \\
      & sd89-ds100 & 10.4 & (9.4) & 10.4 & (7.9) \\
      & sd99-ds90  & 12.6 &(11.1) & 11.4 &(10.8) \\
      & sd99-ds98  & 11.0 & (9.0) &  8.8 & (8.6) \\
\hline                                                 
HSPC  & ds89-sd100 &  4.0 & (3.3) &      &       \\
      & ds99-sd90  &  5.8 & (4.7) &      &       \\
      & ds99-sd98  &  6.3 & (5.0) &      &       \\
      & sd89-ds100 &  5.3 & (4.0) &      &       \\
      & sd99-ds90  &  7.0 & (6.1) &      &       \\
      & sd99-ds98  &  7.8 & (6.7) &      &       \\
\hline                                                 
LSPC  & ds89-sd100 & 10.5 & (8.3) &  7.3 & (5.0) \\
      & ds99-sd90  & 20.4 &(15.2) & 11.4 &(13.2) \\
      & ds99-sd98  & 31.2 &(29.9) &  9.6 & (6.7) \\
      & sd89-ds100 & 11.3 &(10.4) &  8.3 & (6.2) \\
      & sd99-ds90  & 17.0 & (9.3) &  6.0 & (3.9) \\
      & sd99-ds98  & 18.1 &(12.5) &  7.6 & (5.1) \\
\hline
      &            &      &       &      &  (ns) \\
\end{tabular}
\end{table}

\begin{figure}
  \includegraphics[width=170mm]{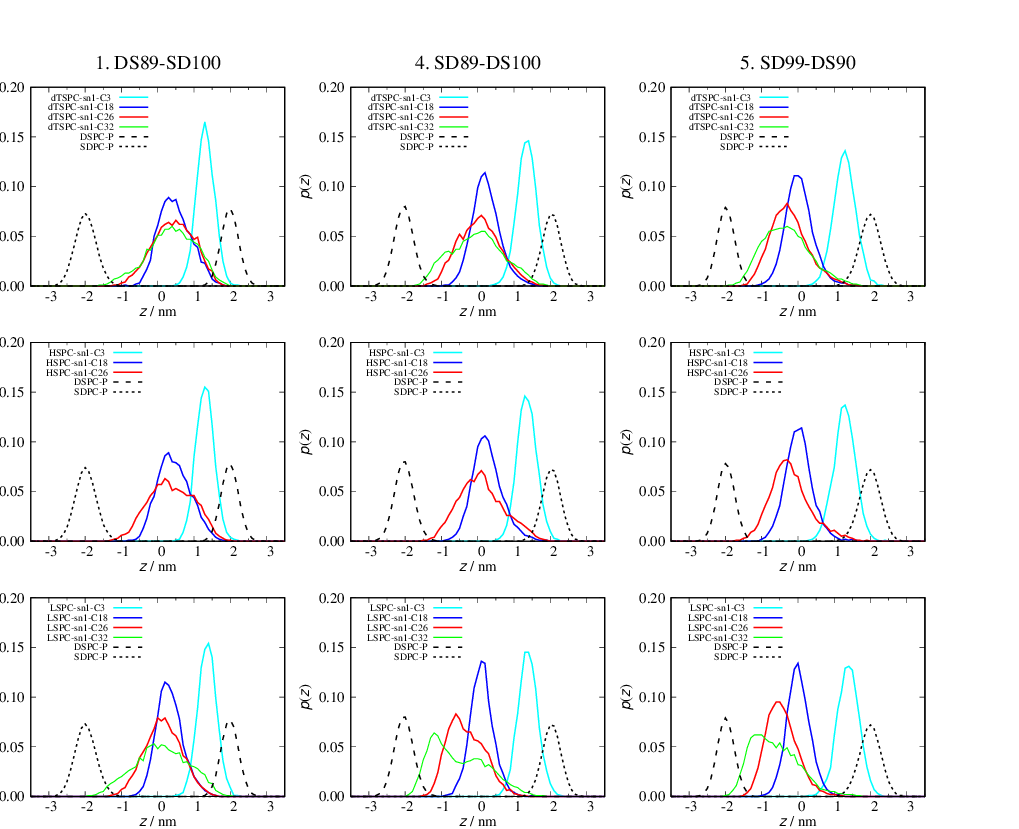}
  \caption{Probability distribution of atoms of dTSPC (top panels), HSPC (middle panels), and LSPC (bottom panels) in the vertical ($z$) direction. 
    The upper and lower leaflets of host membranes consist of SDs or DCs.
          Left to right: DS89-SD100, SD89-DS100, and SD99-DS90.
          The solid lines represent the C$_{3}$, C$_{18}$, C$_{26}$, and C$_{32}$ atoms in the sn-1 chain of the ULCFAs.
          The dashed and dotted lines represent phosphate (P) atoms of DSPC and SDPC, respectively. }
  \label{fig3}
\end{figure}

To analyze the conformation of the ULCFAs,
the distributions of the vertical ($z$) positions of different atoms are plotted in \ref{fig3} and S1:
 C$_{3}$, C$_{18}$, C$_{26}$, and C$_{32}$ atoms in the sn-1 chain of ULCFAs
 and phosphate (P) atoms of DSPC and SDPC (see \ref{fig1} and \ref{fig2}).
C$_{32}$ is the terminal carbon of the dTSPC and LSPC,
 whereas C$_{26}$ is the terminal carbon of the HSPC.
The terminal carbons C$_{32}$ of dTSPC (top panels in \ref{fig3})
 and LSPC (bottom panels in \ref{fig3}) and C$_{26}$ of HSPC (middle panels in \ref{fig3} are located over a broad range,
from $z = -$2 to 2 nm, corresponding to the conformational fluctuations shown in \ref{fig2}.
The $z$-position of P showed a similar distribution in all cases,
 although the lipid molecules and lipid density were different.
These results indicate that the membrane thickness does not depend on the type of host lipid
 or the lipid density difference between the two leaflets under the simulated conditions.

The peaks of the C$_{18}$ and C$_{32}$ positions in \ref{fig3} and Figure S1 are located
 around the middle of the bilayer ($z=0$) but
are slightly shifted to the lower or upper leaflet depending on the simulation conditions.
Thus, the sn-1 chain most frequently adopted an L-shaped conformation.
As the lipid density of the lower leaflet became smaller than that of the upper leaflet,
the  distribution of C$_{32}$  shifted toward the lower leaflet to reduce the difference in lipid density.
We previously reported the density dependence of dTSPC in single-component host membranes~\cite{kk20}.
Here, we confirmed that this dependence is maintained for other ULCFAs (HSPC and LSPC),
 as well as in membranes having asymmetric compositions.
Moreover, \ref{fig3} shows that the sn-1 chain of LSPC is more frequent in the lower leaflet than that of dTSPC,
whereas the C$_{26}$ positions of HSPC and dTSPC exhibit similar distributions.
Therefore, the saturated ultra-long chains exhibited more elongated conformations than the unsaturated chains.
In addition, the truncation of the ultra-long chain had little effect on the conformation of the remaining chains.

\begin{figure}
  \includegraphics[width=170mm]{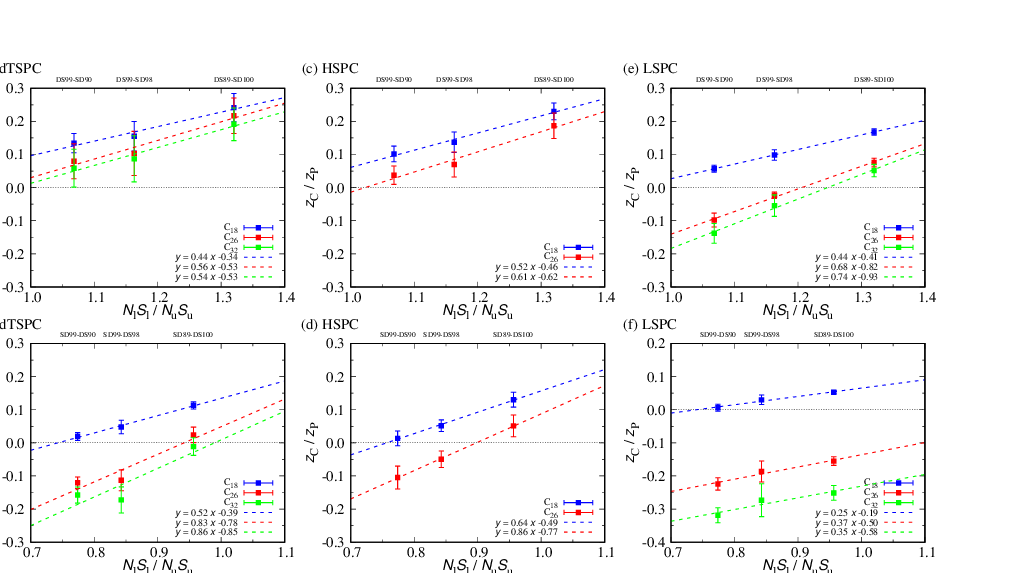}
  \caption{Correlation between the normalized $z$-position of C$_{18}$, C$_{26}$, and C$_{32}$ atoms
            of dTSPC (a and b), HSPC (c and d), and LSPC (e and f) and the lipid-density ratio.
           Top panels (a, c, and e) represent the results for the membrane
            consisting of DSPC and SDPC in the upper and lower leaflet, respectively.
           Bottom panels (b, d, and f) represent the results for the membrane
            consisting of SDPC and DSPC in the upper and lower leaflet, respectively.
           Error bars are calculated from the standard deviations of eight time-windows in single simulation runs.
           The dashed lines were obtained by least-squares fitting.}
  \label{fig4}
\end{figure}

To analyze the effects of lipid density differences on the sn-1 chain of the ULCFAs further,
 we calculated the mean $z$-position of the C atoms normalized by the $z$-position of the P atoms in the sn-1 chain
 as a function of the lipid density ratio, as shown in \ref{fig4}.
The lipid density ratio is defined as $N_{\rm l}S_{\rm l}/N_{\rm u}S_{\rm u}$,
 where $N_{\rm l}$ and $N_{\rm u}$ are the numbers of lipid molecules in the lower and upper leaflets, respectively,
 except for ULCFA.
$S_{\rm l}$ and $S_{\rm u}$ represent the area per lipid of DSPC or SDPC in the lower and upper leaflets, respectively.
A linear correlation was obtained between the $z$-positions of carbon atoms and the lipid-density differences,
 as shown in \ref{fig4}, in all cases, as in our previous study~\cite{kk20}. 
The sn-1 chain of the LSPC is embedded more deeply in the opposite leaflet (compare the left and right panels in \ref{fig4}). 

The sn-1 chain quickly changes these three conformations under the influence of thermal fluctuations, as shown in Figure S2.
The transit time between the elongated and turned conformations was calculated as
the time at which the $z$-position of C$_{32}$ or C$_{26}$ moved between the upper and lower boundaries.
The upper and lower boundaries are defined as $z = 1.5$ and $-1.5$ nm for C$_{32}$
and $z = 1.2$ and $-1.2$ nm for C$_{26}$, respectively.
The average transit time were approximately 10 ns, as shown in \ref{t3}.
The transit times of C$_{26}$ of the LSPC were $10$--$30$ ns, longer than those of the others.
This is because C$_{26}$ of LSPC does not overcome the upper or lower boundary
owing to the larger peaks in the $z$ distribution of C$_{26}$ of LSPC (see the bottom panels in \ref{fig3}).

\subsubsection{Order Parameters}
\begin{figure}
  \includegraphics[width=80mm]{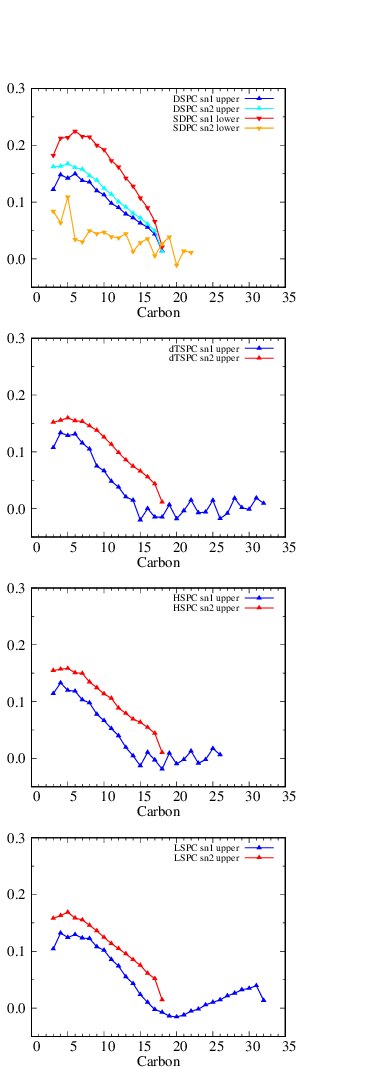}
  \caption{Order parameter profiles, $-S_{\rm CD}$, of (a) DSPC and SDPC, (b) dTSPC, (c) HSPC, and (d) LSPC for DS89-SD100.}
  \label{fig5}
\end{figure}

To investigate the orientation of the acyl chains of host lipids and ULCFAs,
 the lipid order parameters, $S_{\rm CD}$, were calculated by

\begin{equation}
S_{\rm CD} = \langle \frac{3 \cos ^2 \alpha -1}{2} \rangle, \label{eq1}
\end{equation}
Here $\alpha$ is the angle between C-H bond vector and the bilayer normal.
The brackets indicate the average over time and the lipid molecules.
\ref{fig5} shows the calculated $-S_{\rm CD}$ values for the host lipids and ULCFAs for DS89-SD100.
These profiles were not sensitive to the lipid ratio in the two leaflets. 
The order profiles of host lipids and dTSPC were similar to those in our previous study \cite{kk20}.

The order of host lipids is shown in \ref{fig5}(a).
For DSPC,
the overall order profile of the sn-1 chain is similar to that of the sn-2 chain,
 because the two acyl chains of DSPC are equivalent.
For SDPC,
 the order profile of the sn-1 chain is higher than that of the sn-2 chain;
 therefore the sn-1 chain is more disordered than the sn-2 chain, because the sn-2 chain has six double bonds.

The order profiles of the dTSPC, HSPC, and LSPC bilayers are shown in \ref{fig5}(b), (c), and (d), respectively.
The order of the sn-1 chain decreased to 0 at C$_{15}$ and
 exhibited a low order in the longer region until the terminal (C$_{32}$).
The order of the sn-2 chain is similar to that of the acyl chains in DSPC, as shown in \ref{fig5}(a),
 because the sn-2 chain is also equivalent to the acyl chains in DSPC.

In the case of HSPC,
 the order profile of the sn-1 chain until C$_{26}$ exhibits a similar dependence to that of dTSPC
because the sn-1 chain in HSPC is equivalent to the sn-1 chain until C$_{26}$ in dTSPC.
This is consistent with the slight difference in the $z$-position of C$_{26}$ between dTSPC and HSPC, as shown in \ref{fig3}.
In contrast,
 the order profile of the longer region of the sn-1 chain (C$_{16}$-C$_{32}$) of LSPC exhibits
 different behavior to that of dTSPC (compare \ref{fig5}(b) and \ref{fig5}(d)).
The sn-1 chain of the LSPC has no double bonds, and
 the order profile becomes higher toward the terminal.
This agrees with the shift of the $z$-position of C$_{32}$ between LSPC shown in \ref{fig3} and \ref{fig4}, 
 which indicates a more elongated conformation. 
The order of the sn-2 chains of HSPC and LSPC was similar behavior to that of the sn-2 chain of dTSPC.

\subsection{dTSPC with Cholesterol}
\begin{figure}
  \includegraphics[width=80mm]{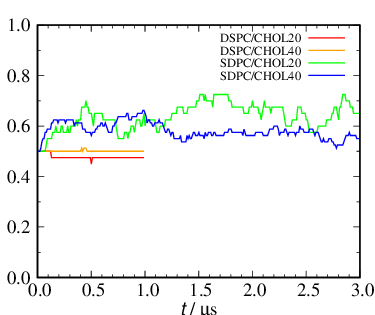}
  \caption{Change in fraction of cholesterol in the upper leaflet with time.}
  \label{fig6}
\end{figure}

We investigated the interactions between the dTSPC and cholesterol in asymmetric bilayers.
Cholesterol molecules were embedded in the asymmetric bilayers with number fractions of 20\% or 40\%, as shown in \ref{t2}.
In the initial state of each simulation,
the fraction of cholesterol in the upper leaflet ($N_{\rm upper}/N_{\rm total}$) was set to 0.5,
that is, 20 and 40 cholesterol molecules per leaflet for CHOL20 and CHOL40, respectively.
\ref{fig6} shows the change in the cholesterol fraction in the upper leaflet with time.
In the case of the DSPC bilayers, the fraction is almost constant at 1 $\mu$s,
showing that the flip--flop time of cholesterol is much longer than 1 $\mu$s.
Conversely, in the case of the SDPC bilayers,
 the fraction frequently changed during 3 $\mu$s,
 such that the density difference between the two leaflets was reduced by cholesterol.
The flip--flop times per cholesterol molecule were 0.52 and 0.66 $\mu$s for SDPC/CHOL20 and SDPC/CHOL40, respectively.
This is much faster than those of phospholipids (hours or days)
but slower than those of ULCFAs.
A similar time scale of the cholesterol flip--flop was previously reported
 in the MD simulations of other lipid compositions \cite{jo10,ogus12,gu19}.

\begin{figure}
  \includegraphics[width=80mm]{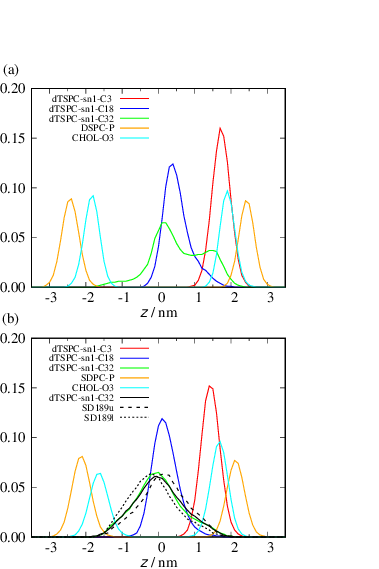}
  \caption{Probability distribution of atoms for (a) DSPC/CHOL and (b) SDPC/CHOL.
The solid lines represent the data for DSPC/CHOL40 and SDPC/CHOL40.
The solid black line in (b) represents C32 for SDPC/CHOL20.
The dashed and dotted lines represent C32 in the absence of cholesterol molecules for SD189u and SD189l obtained from our previous study \cite{kk20}, respectively.
The membrane of SD189u has 89 SDPCs and 100 SDPCs in the upper and lower leaflets, respectively,
 whereas that of SD189l has 99 SDPCs and 90 SDPCs, respectively.
}
  \label{fig7}
\end{figure}

We observed no specific, direct interaction between dTSPC and cholesterol molecules from snapshots of the simulations.
However, they indirectly interact via the difference in lipid density between the two leaflets.
\ref{fig7} shows the probability distribution of the C$_3$, C$_{18}$, and C$_{32}$ atoms of the sn-1 chain of the dTSPC,
 P of the host lipid, and O3 of cholesterol.
The distribution of the terminal atom (C$_{32}$) of dTSPC has two peaks
 around $z_{\rm C} = 0.0$ and 1.5 nm (the normalized average position $z_{\rm C}/z_{\rm p}= 0.23$) 
in the case of DSPC/CHOL40 shown in \ref{fig7}(a),
 whereas the distribution of C$_{32}$ has a single peak around $z_{\rm C} = 0.0$ nm
 in the case of SDPC/CHOL40 shown in \ref{fig7}(b).
The dTSPC distribution in DSPC/CHOL40 is similar to that of the same number ratio of DSPCs (89 DSPCs and 100 DSPCs
 in the upper and lower leaflets, respectively) in the absence of cholesterol observed in our previous study~\cite{kk20}. 
However, the dTSPC distribution in SDPC/CHOL40 is different from that for the same number ratio of SDPCs
 without cholesterol molecules (SD189u) obtained in our previous study \cite{kk20} (compare dashed and solid lines in \ref{fig7}(b)).
The C32 of the dTSPC is distributed at the center of the bilayer and
is between the dashed line (SD189u) and dotted line (SD189l: 99 SDPCs and 90 SDPCs in the upper and lower leaflet, respectively).
This is due to the reduction in the density difference resulting from the  flip--flop of cholesterol.
The dTSPC behaves as in an SDPC membrane with no density difference between the two leaflets.

\subsection{Conclusion}
In this study,
 we have performed MD simulations for three types of ULCFAs (dTSPC, HSPC, and LSPC)
 embedded in DSPC/SDPC asymmetric bilayer membranes.
All types of ULCFAs exhibit a large conformational change in the ultra-long sn-1 chain under thermal fluctuations.
The ultra-long chain fluctuates from the upper to the lower leaflets
 and takes elongated, L-shaped, and turned conformations.
As the lipid density of the opposite leaflet decreases,
 the ratio of the elongated conformation increases, reducing the lipid density differences between the two leaflets.
We have observed linear correlations between the position of the terminal atoms
 and lipid density differences in all simulated conditions,
 as obtained in our previous study~\cite{kk20} for single-component host membranes.
These behaviors are conserved for all types of ULCFAs under all simulated conditions.
Thus, we concluded that these are general properties of ULCFAs.

When the ultra-long chain of the dTSPC is replaced with a saturated chain of the same length (LSPC),
 an elongated conformation is more often formed.
Interestingly, when the long chain of the dTSPC is truncated at C26 (HSPC),
 the conformation of the remaining ULCFA part is unchanged.
These ULCFAs (ultra-long chain with saturation and medium length) are 
 hardly observed in phospholipids in living cells.
However, we found no drastic differences in the physicochemical properties,
 although quantitative differences are observed.
Thus, the selection of ULCFAs in cells is likely determined by biological factors
such as interactions with specific proteins and responses to biological signals.
Although this study indicates the role of ULCFAs as membrane components,
 the possible role of ULCFAs as a precursor of lipid mediators requires further examination.

The lipid density difference in the SDPC membrane is reduced by the flip--flop of cholesterol molecules
but this flip--flop is slower than the conformational change of ULCFAs.
However, after the density difference was compensated by  cholesterol molecules,
 the ULCFA conformation returns to that in the no-density-difference condition.
Thus, we consider that ULCFAs initially respond to the density difference and,
later, cholesterol removes the difference in biomembranes.

\begin{acknowledgement}
This research was supported by JSPS KAKENHI JP21K03481 (H.N.),
 the Japan Agency for Medical Research and Development (AMED)-CREST 22gm0910011 (H.S.),
  and the AMED Program for Basic and Clinical Research on Hepatitis 22fk0210091 (H.S.).
MD simulations were carried out by using the facilities of the Supercomputer Center, 
 Institute for Solid State Physics, University of Tokyo.
\end{acknowledgement}

\begin{suppinfo}
\end{suppinfo}

\bibliography{ulcfa-jpcb}

\end{document}